\documentclass{sigchi-ext}
\usepackage[T1]{fontenc}
\usepackage{textcomp}
\usepackage[scaled=.92]{helvet} 
\usepackage{graphicx} 
\usepackage{balance}  
\usepackage{booktabs} 
\usepackage{ccicons}  
\usepackage{ragged2e} 
\usepackage{caption}

\usepackage{marginnote} 
\usepackage{paralist}

\def\plaintitle{Fair and Responsible AI: A Focus on the Ability to Contest} 
\def\emptyauthor{}
\def\plainkeywords{Contestability; explainability; algorithmic fairness, ethics.}

\title{Fair and Responsible AI: A Focus on the Ability to Contest}

\numberofauthors{3}
\author{%
  \alignauthor{%
    \textbf{Henrietta Lyons}\\
    \affaddr{University of Melbourne} \\
    \affaddr{Parkville, VIC 3010, AUS} \\
    \email{hlyons@student.unimelb.edu.au}
   }\alignauthor{%
    \textbf{Tim Miller}\\
    \affaddr{University of Melbourne} \\
    \affaddr{Parkville, VIC 3010, AUS} \\
    \email{tmiller@unimelb.edu.au}} \alignauthor{%
    \vspace{10pt} \textbf{Eduardo Velloso}\\
      \affaddr{University of Melbourne} \\
    \affaddr{Parkville, VIC 3010, AUS} \\
    \email{eduardo.velloso@unimelb.edu.au} }\alignauthor{%
     }\alignauthor{%
  }}

\definecolor{linkColor}{RGB}{6,125,233}
\hypersetup{%
  pdftitle={\plaintitle},
  pdfauthor={\emptyauthor},
  pdfkeywords={\plainkeywords},
  bookmarksnumbered,
  pdfstartview={FitH},
  colorlinks,
  citecolor=black,
  filecolor=black,
  linkcolor=black,
  urlcolor=linkColor,
  breaklinks=true,
}

\begin{document}

\CopyrightYear{2020}
\setcopyright{rightsretained}
\conferenceinfo{Fair and Responsible AI Workshop (CHI'20)}{April  25--30, 2020, Honolulu, HI, USA}
\isbn{978-1-4503-6819-3/20/04}
\doi{https://doi.org/10.1145/3334480.XXXXXXX}
\copyrightinfo{\acmcopyright}

\maketitle

\RaggedRight{} 

\begin{abstract}
As the use of artificial intelligence (AI) in high-stakes decision-making increases, the ability to contest such decisions is being recognised in AI ethics guidelines as an important safeguard for individuals. Yet, there is little guidance on how AI systems can be designed to support contestation. In this paper we explain that the design of a contestation process is important due to its impact on perceptions of fairness and satisfaction. We also consider design challenges, including a lack of transparency as well as the numerous design options that decision-making entities will be faced with. We argue for a human-centred approach to designing for contestability to ensure that the needs of decision subjects, and the community, are met.
\end{abstract}



\section{Introduction}

 There is great potential for Artificial Intelligence (AI) to enhance decision-making, by making it more accurate, efficient, and scalable than human decision-making \cite{Binns18, Lee19}. To harness these benefits, AI systems should be designed responsibly, to ensure that they are fair, accountable, and transparent \cite{Dignum}. This is particularly important given the increasing use of AI in high-stakes decision-making, including sentencing, hiring, and loan application determination \cite{Lee19}. 
 
 In response to calls for AI systems to be designed, developed, and deployed responsibly, numerous AI ethics guidelines have been produced. One `safeguard' that is gaining traction within these guidelines is the ability to contest AI decisions (see sidebar for examples) \cite{Jobin19}. Article 22(3) of the European Union's General Data Protection Regulation provides a legal ‘right to contest’ decisions made using solely automated processes. However, none of these documents provide guidance on how AI systems should be designed to enable contestation. In this paper, we outline why design is important, what the design challenges are, and our human-centred approach to designing for contestability.

\marginpar{%
  \vspace{-170pt} \fbox{%
    \begin{minipage}{0.925\marginparwidth}
      \textbf{Examples of Ethical AI Guidelines Calling for the Ability to Contest} \\
      \vspace{1pc} Ethics Guidelines for Trustworthy AI (High-Level Expert Group on Artificial Intelligence): 
      \textit{"[T]here are many different interpretations of fairness, we believe that fairness has both a substantive and a procedural dimension... The procedural dimension of fairness entails the ability to contest and seek effective redress against decisions made by AI systems and by the humans operating them."} \cite{EUGuide}  \\
      \vspace{1pc} Principles on Artificial Intelligence (OECD):
      \textit{"There should be transparency and responsible disclosure around AI systems to ensure that people understand AI-based outcomes and can challenge them."} \cite{OECD}  \\
      \vspace{1pc} AI Ethics Framework (Australia):
      \textit{(Principle 7) "Contestability: When an AI system significantly impacts a person, community, group or environment, there should be a timely process to allow people to challenge the use or output of the AI system."} \cite{AUAI}\\
    \end{minipage}}}

\section{The importance of design}
The importance of designing AI systems to enable `contestability' has been acknowledged in HCI and Algorithmic Accountability literature (e.g. \cite{Hirsch17, Almada19}). Using a legal lens, the Algorithmic Accountability work has taken a theoretical approach to proposing requirements of a contestation scheme \cite{Bayamlioglu}, and design requirements that enable contestation \cite{Almada19}. Within HCI, the focus of contestability research has been on the ability of expert users to work interactively with a system to contest its output \cite{Hirsch17}. 

HCI researchers \cite{Lee19, Binns18, Grgic18} have also drawn on organisational psychology literature to study how the design of AI systems used in decision-making impact human perceptions of procedural fairness, `procedural justice' \cite{Thibaut}. The procedural justice literature indicates that having a legitimate way to contest a decision increases a person's perception of procedural fairness, which can impact their perception of the fairness of the decision (`distributive justice'), their choice to accept or contest a decision, and their attitude towards the entity making the decision \cite{Thibaut, Leventhal}. In line with this literature \cite{Leventhal}, Lee et al \cite{Lee19} found that having `outcome control', the ability to correct or appeal a decision, in a cooperative group allocation task improved participants' perceptions of the fairness of the outcome. 

The procedural justice literature also indicates that design of a contestation process (not just its availability) impacts perceptions of procedural fairness. For example, having the same decision-maker assess the decision on appeal is seen as less fair than having a new decision-maker \cite{Leventhal}. In addition, in a study of content moderation across social media platforms, Myers West \cite{MyersWest2018} found that users were dissatisfied with contestation processes, reporting a lack of clear instruction about how to lodge an appeal, receiving no reply or resolution having lodged a challenge, and no access to human intervention. These findings indicate that the design of a contestation process matters.

\section{Design challenges}
 To meaningfully challenge a decision, a decision subject requires some form of information in order to understand the decision, decide whether to contest, and to use as grounds for contestation. Many AI systems used in decision-making are effectively ``black boxes" \cite{Rudin18}; their decision-making processes are hidden due to the use of complex algorithms or techniques (e.g. deep learning) or intentionally by companies to protect trade secrets \cite{Burrell}. This opacity makes it difficult to understand why a decision was made, and consequently, to contest it in any meaningful way. In contrast, with human decision-making a person can generally seek an explanation from the decision maker as to why a decision was made. Often, in high-stakes decisions, reasons must be documented during the decision-making process to mitigate the issue of an inaccurate post-hoc explanation. Promisingly, the field of explainable artificial intelligence (XAI) is progressing work into explainability \cite{Watcher18}. To date, XAI has not focused on providing explanations for contestation specifically, which offers a new avenue of research. 
 
 A second design challenge is that there are many ways to contest a decision \cite{Leventhal}. For example, existing contestation processes for human decisions (e.g. internal review, complaints mechanisms, external review via tribunal or court) could be adapted for decisions made using AI. However, with decisions made at scale, leaving appeal processes to a court to determine would overwhelm an already pressured system. Low perceptions of fairness are associated with procedures that are time consuming, costly and resource intensive \cite{Leventhal}. An alternative contestation process might involve a decision subject directly contesting a decision within AI system via an interface. However, the novelty of this approach coupled with a lack of human touch could negatively impact perceptions of fairness. With an abundance of design choices, it is difficult to know where to begin.
 
 \marginpar{%
  \vspace{-160pt} \fbox{%
    \begin{minipage}{0.925\marginparwidth}
      \textbf{Sample of preliminary findings from our current research} \\
      \vspace{1pc}
      AI systems are not isolated, but exist in socio-technical contexts with existing legal frameworks, political systems, and social norms, that need to be considered when designing for contestation  \\
         \vspace{1pc} 
      Different processes for contestation are likely to be required depending on the context in which a decision is being made \\
        \vspace{1pc} 
      Contestation processes need to be clear and easy to access \\
      \vspace{1pc} 
      Contestation processes should align with human rights, to ensure equality, be designed for accessibility, and to provide compensation \\
      \vspace{1pc}
      Lack of transparency is an issue; explainability is important \\
    \end{minipage}}\label{sec:sidebar2} }
 
 We suggest that taking a human-centred approach to explore how people conceptualise contestability in relation to AI systems is a key first step in designing for meaningful contestation. To understand the needs of decision subjects, and the expectations of the community more generally, we are currently conducting a thematic analysis of submissions made to Australia's ‘Artificial Intelligence: Australia’s Ethics Framework’, a discussion paper that proposed `contestability' as a core ethical principle \cite{AUAI}. The sidebar contains a sample of our preliminary findings.

\section{Conclusion}
 The increasing use of AI in high-stakes decision-making, which has been deployed without appropriate safeguards like procedural fairness, has had a significant, negative impact on thousands of people, from teachers losing their jobs \cite{Houston} to the erroneous loss of medical benefits \cite{Eubanks}. To reduce negative consequences, AI systems must be responsibly designed, developed, and deployed \cite{Dignum}. Though the ability to contest decisions is not the only mechanism required to ensure that AI systems are `fair', it is a crucial safeguard, and in some circumstances, a legal requirement. How access to contestation, and the contestation process itself, is designed is important given the impact on perceptions of fairness and satisfaction. Yet, there are many design challenges including opacity, and an abundance of design options. A key first step in designing for meaningful contestation is to explore and understand the needs of decision subjects as well as the community more generally.


\section{Acknowledgements}
Henrietta Lyons is supported by the Melbourne School of Engineering Ingenium scholarship program. This research was partly funded by Australian Research Council Discovery Grant DP190103414 \emph{Explanation in Artificial Intelligence: A Human-Centred Approach}. Eduardo Velloso is the recipient of an Australian Research Council Discovery
Early Career Researcher Award (Project Number: DE180100315) funded by the Australian Government.

\balance{} 

\bibliographystyle{SIGCHI-Reference-Format}
\bibliography{sample.bib}


\begin{thebibliography}{00}


\ifx \showCODEN    \undefined \def \showCODEN     #1{\unskip}     \fi
\ifx \showDOI      \undefined \def \showDOI       #1{{\tt DOI:}\penalty0{#1}\ }
  \fi
\ifx \showISBNx    \undefined \def \showISBNx     #1{\unskip}     \fi
\ifx \showISBNxiii \undefined \def \showISBNxiii  #1{\unskip}     \fi
\ifx \showISSN     \undefined \def \showISSN      #1{\unskip}     \fi
\ifx \showLCCN     \undefined \def \showLCCN      #1{\unskip}     \fi
\ifx \shownote     \undefined \def \shownote      #1{#1}          \fi
\ifx \showarticletitle \undefined \def \showarticletitle #1{#1}   \fi
\ifx \showURL      \undefined \def \showURL       #1{#1}          \fi

\bibitem{Houston}
{\em \textup{Houston {F}ederation of {T}eachers, {L}ocal 2415, et al v
  {H}ouston {I}ndependent {S}chool {D}istrict, 251 {F}.{S}upp.3d 116} (2017)}.
\newblock


\bibitem{Almada19}
{Marco Almada}. 2019.
\newblock \showarticletitle{Human intervention in automated decision-making:
  Toward the construction of contestable systems}. In {\em Proceedings of the
  Seventeenth International Conference on Artificial Intelligence and Law}.
  2--11.
\newblock


\bibitem{Bayamlioglu}
{Emre Bayamlioglu}. 2018.
\newblock \showarticletitle{Contesting Automated Decisions}.
\newblock {\em European Data Protection Law Review\/}  {4} (2018), 433--446.
\newblock


\bibitem{Binns18}
{Reuben Binns}, {Max Van~Kleek}, {Michael Veale}, {Ulrik Lyngs}, {Jun Zhao},
  {and} {Nigel Shadbolt}. 2018.
\newblock \showarticletitle{It's Reducing a Human Being to a Percentage}.
\newblock {\em Proc of the 2018 CHI Conference on Human Factors in Computing
  Systems - CHI '18\/} (2018).
\newblock
\showISBNx{9781450356206}


\bibitem{Burrell}
{Jenna Burrell}. 2016.
\newblock \showarticletitle{How the machine ‘thinks’: Understanding opacity
  in machine learning algorithms}.
\newblock {\em Big Data and Society\/} (2016), 1--12.
\newblock


\bibitem{Dignum}
{Virginia Dignum}. 2019.
\newblock {\em {R}esponsible {A}rtificial {I}ntelligence: How to Develop and
  Use AI in a Responsible Way}.
\newblock Springer.
\newblock


\bibitem{Eubanks}
{Virginia Eubanks}. 2018.
\newblock {\em Automating Inequality: How High-Tech Tools Profile, Police and
  Punish the Poor}.
\newblock St Martin's Publishing Group, Hillsdale, NJ.
\newblock


\bibitem{OECD}
{Organisation for Economic Co-operation} {and} {Development}. 2019.
\newblock OECD Principles on Artificial Intelligence.
\newblock   (2019).
\newblock
\newblock
\shownote{Retrieved 30 January 2020 from
  https://www.oecd.org/going-digital/ai/principles/.}


\bibitem{Grgic18}
{Nina Grgic-Hlaca}, {Muhammad~Bilal Zafar}, {Krishna~P. Gummadi}, {and} {Adrian
  Weller}. 2018.
\newblock \showarticletitle{Beyond Distributive Fairness in Algorithmic
  Decision Making: Feature Selection for Procedurally Fair Learning}. In {\em
  Proc of Thirty-Second AAAI Conference on Artificial Intelligence} {\em
  (AAAI-18)}. 51--60.
\newblock


\bibitem{Hirsch17}
{Tad Hirsch}, {Kritzia Merced}, {Shrikanth Narayanan}, {Zac~E Imel}, {and}
  {David~C Atkins}. 2017.
\newblock \showarticletitle{Designing contestability: Interaction design,
  machine learning, and mental health}. In {\em Proceedings of the 2017
  Conference on Designing Interactive Systems}. 95--99.
\newblock


\bibitem{Jobin19}
{Anna Jobin}, {Marcello Ienca}, {and} {Effy Vayena}. 2019.
\newblock \showarticletitle{The global landscape of AI ethics guidelines}.
\newblock {\em Nature Machine Intelligence\/} 1 (2019), 389–399.
\newblock


\bibitem{Lee19}
{Min~Kyung Lee}, {Anuraag Jain}, {Hea~Jin Cha}, {Shashank Ojha}, {and} {Daniel
  Kusbit}. 2019.
\newblock \showarticletitle{Procedural justice in algorithmic fairness:
  Leveraging transparency and outcome control for fair algorithmic mediation}.
\newblock {\em Proceedings of the ACM on Human-Computer Interaction\/} {3},
  CSCW (2019), 1--26.
\newblock


\bibitem{Leventhal}
{Gerald~S Leventhal}. 1980.
\newblock \showarticletitle{What should be done with equity theory?}
\newblock In {\em Social exchange}. Springer, 27--55.
\newblock


\bibitem{MyersWest2018}
{Sarah Myers~West}. 2018.
\newblock \showarticletitle{Censored, suspended, shadowbanned: User
  interpretations of content moderation on social media platforms}.
\newblock {\em New Media {\&} Society\/} {2}, 11 (2018), 4366--4383.
\newblock


\bibitem{AUAI}
{Australian Government~Department of Industry~Innovation} {and} {Science}.
  2019.
\newblock AI Ethics Framework.
\newblock   (2019).
\newblock
\newblock
\shownote{Retrieved 30 January 2020 from
  https://www.industry.gov.au/data-and-publications/building-australias-artificial-intelligence-capability/ai-ethics-framework.}


\bibitem{EUGuide}
{Independent High-Level Expert~Group on Artificial~Intelligence}. 2019.
\newblock Ethics Guidelines for Trustworthy AI.
\newblock   (2019).
\newblock
\newblock
\shownote{Retrieved on January 30, 2020 from
  https://ec.europa.eu/digital-single-market/en/news/ethics-guidelines-trustworthy-ai.}


\bibitem{Rudin18}
{Cynthia Rudin}. 2018.
\newblock Stop Explaining Black Box Machine Learning Models for High Stakes
  Decisions and Use Interpretable Models Instead.
\newblock   (2018).
\newblock


\bibitem{Thibaut}
{John Thibaut} {and} {Laurens Walker}. 1975.
\newblock {\em Procedural Justice: A Psychological Analysis}.
\newblock Lawrence Erlbaum Associates, Hillsdale, NJ.
\newblock


\bibitem{Watcher18}
{Sandra Wachter}, {Brent Mittelstadt}, {and} {Chris Russell}. 2018.
\newblock \showarticletitle{Counterfactual Explanations without Opening the
  Black Box: Automated Decisions and the GDPR}.
\newblock {\em Harvard Journal of Law and Technology\/} {31}, 2 (2018),
  841--887.
\newblock


\end{thebibliography}

\end{document}